\documentclass[prl,preprintnumbers,amsmath,amssymb,superscriptaddress,preprint]{revtex4-2}
\usepackage{graphicx}
\usepackage{dcolumn}
\usepackage{bm}
\usepackage{color}
\usepackage{amsmath}
\usepackage{amssymb}
\usepackage{latexsym}
\usepackage{multirow}
\usepackage{xcolor}
\usepackage{hyperref}
\usepackage[capitalise]{cleveref}
\usepackage{textcomp}
\usepackage{soul}
\begin{document}
\title{Measurement of the depth-dependent local dynamics in thin polymer films through rejuvenation of ultrastable  glasses} 
\author{Saba Karimi}
\affiliation{Department of Physics and Astronomy, University of Waterloo, Waterloo, Ontario, Canada.}
\author{Junjie Yin}
\affiliation{Department of Physics and Astronomy, University of Waterloo, Waterloo, Ontario, Canada.}
\author{Thomas Salez}
\affiliation{Univ. Bordeaux, CNRS, LOMA, UMR 5798, F-33400, Talence, France.}
\author{James A. Forrest} 
\thanks{corresponding author email: jforrest@uwaterloo.ca}
\affiliation{Department of Physics and Astronomy, University of Waterloo, Waterloo, Ontario, Canada.}
\affiliation{Univ. Bordeaux, CNRS, LOMA, UMR 5798, F-33400, Talence, France.}
\date{\today}
\begin{abstract}
We measure the isothermal rejuvenation of stable glass films of poly(styrene) and poly(methylmethacrylate). We demonstrate that the propagation of the front responsible for the transformation to a supercooled-liquid state can serve as a highly localized probe of the local supercooled dynamics.  We use this connection to probe the depth-dependent relaxation rate with nanometric precision for a series of polystyrene films over a range of temperatures near the bulk glass transition temperature.  The analysis shows the spatial extent of enhanced surface mobility and reveals the existence of an unexpected large dynamical length scale in the system. The results are compared with the cooperative-string model for glassy dynamics.  The data reveals that the film-thickness dependence of whole film properties arises only from the volume fraction of the near-surface region.  While the dynamics at the middle of the samples shows the expected bulk-like temperature dependence, the near-surface region shows very little dependence on temperature. 
\end{abstract}
\maketitle

Experimentally measuring the local dynamics in  glasses and supercooled liquids is key to understanding the glass transition, and remains an ongoing quest. As an illustration, since the first report of anomalous values of the glass-transition temperature $T_{\textrm{g}}$ in thin polystyrene  films ~\cite{keddie1994size}, it has been an overarching, and so far unattainable, objective to measure the depth-dependent dynamics in thin films. Many research groups have measured $T_{\textrm{g}}$ in thin polymer films, primarily polystyrene (PS),  and that body of work has been summarized in a number of review articles~\cite{roth2005glass,forrest2001glass,ediger2014dynamics,alcoutlabi2005effects}.  A lack of complete consensus still exists, even for this single material (PS). This is likely due to differences in how the $T_{\textrm{g}}$ values are measured and to the fact that $T_{\textrm{g}}$ is inherently an indirect measure of dynamics. Simulations~\cite{fujimoto2020energy,baljon2010simulated,hao2021mobility,ivancic2020dynamic,torres2000molecular,zhou2017short} and models~\cite{forrest2014does,lipson2010local,salez2015cooperative} are able to make predictions on the local dynamics throughout the film, but a global dilatometric $T_{\textrm{g}}$ needs to be derived from those dynamics for comparison with available experimental measurements. Therefore, it is essential to find ways to experimentally probe the local dynamics directly~\cite{zhang2018we}, and with a nanometric depth resolution.

Direct measurements of the dynamics are much less common than those of $T_{\textrm{g}}$. Film-averaged dynamics have been measured using a variety of techniques~\cite{forrest1998relaxation,svanberg2007glass,yuan2022microscale,o2005rheological}, and indirect measures of the dynamics have also been performed by cooling-rate-dependent measures of $T_{\textrm{g}}$~\cite{fakhraai2005probing,koh2008structural}. Local measures of dynamics have been almost exclusively directed at the free-surface region, as it is the most accessible, and the associated properties have been linked to the origin for the measured $T_{\textrm{g}}$ reductions~\cite{tian2022surface,salez2015cooperative,sharp2003free}. These efforts collectively provide direct evidence for a more mobile layer in the near-surface region of a number of glassy materials~\cite{chai2014direct,daley2012comparing,zhang2017decoupling,li2022surface}, and thus a geometrically-induced dynamical heterogeneity. This heterogeneity in dynamics between the near-surface and bulk regions is only observed for temperatures $T$ below $T_{\textrm{g}}$. Local measures of dynamics throughout the entire film are extremely rare, and uniformly lack the desired depth sensitivity.  Fluorescently-labelled layers have been used to measure both local $T_{\textrm{g}}$ values in PS and structural relaxation in PMMA~\cite{ellison2003distribution,priestley2005structural}. The minimum thickness of these labelled layers was 10 nm. This places severe limits on the depth resolution. In those cases, the sensitivity of a dye molecule to local density is used to infer information about the dynamics. Besides structural relaxation, $\beta$-detected NMR was used with implanted Li ions to specifically probe the local phenyl dynamics  in PS films. In those experiments, there is a depth resolution of a few nanometers near the free surface, but the resolution is much poorer in the interior of the film due to range straggling~\cite{mckenzie2015enhanced,mckenzie2022beta}. Another experiment probed reorientation of dopants in PMMA films with a depth resolution of 12 nm~\cite{oba2012relaxation}.   All of these approaches, while important advances,  require the introduction of a probe impurity and suffer from significant resolution limitations.  There is still no report of structural relaxation throughout a supercooled PS-film sample.

Glasses with high thermodynamic and kinetic stabilities, and high density, were first discovered in 2007~\cite{swallen2007organic}. Since then many other examples have been prepared and studied experimentally~\cite{rodriguez2022ultrastable,ediger2017perspective}, as well as numerically~\cite{parmar2020ultrastable,berthier2020measure,lyubimov2013model,lin2014molecular}. The surface mobile layer in a glass has been directly linked to the ability to prepare these ultrastable glasses by vapour deposition~\cite{berthier2017origin}. Upon heating, thin films of such materials rejuvenate to the normal supercooled liquid by propagation of a transformation front originating from the free surface~\cite{flenner2019front,swallen2009stable} -- a situation reminiscent of many  phase-transition fronts in physics, such as freezing fronts.  The propagation velocity of that front has been correlated to the relaxation dynamics of the supercooled liquid~\cite{rodriguez2022ultrastable}. This connection is usually of the form $v=C\tau_{\alpha}^{-\gamma}$, where $v$ is the rejuvenation-front velocity, $\tau_\alpha$ is the alpha relaxation time, and $\gamma$ is an exponent in the range $\sim 0.8-1$ ~\cite{rodriguez2022ultrastable}. Previous studies~\cite{dalal2015influence,walters2015thermal}  have modelled ellipsometric data to identify the position of the front as a function of time.  Since the front is highly localized in space,  the front-propagation rate should provide a precise measure of the local dynamics at each specific depth as the front moves from the free surface to the supporting substrate of the sample. In this article, we use the established mechanism of rejuvenation by front propagation in thin films as well as the link between rejuvenation-front velocity and material relaxation properties to demonstrate that ellipsometric measurements of isothermal rejuvenation of ultrastable glasses can be used to probe the depth-dependent local dynamics of supercooled thin polymer films at temperatures near the dilatometric $T_{\textrm{g}}$ with nanometric precision throughout the film. We compare the measurements with calculations based on the cooperative-string theory for supercooled dynamics in thin films. The results provide an independent and reasonably direct measure of the size of the surface region, and reveal the existence of an intriguing long-range dynamical effect. In addition the results indicate a very narrow range of temperatures above $T_{\textrm{g}}$ where enhanced surface dynamics are observed. 
\begin{figure}[h!]
\centering
\includegraphics[width=\textwidth]{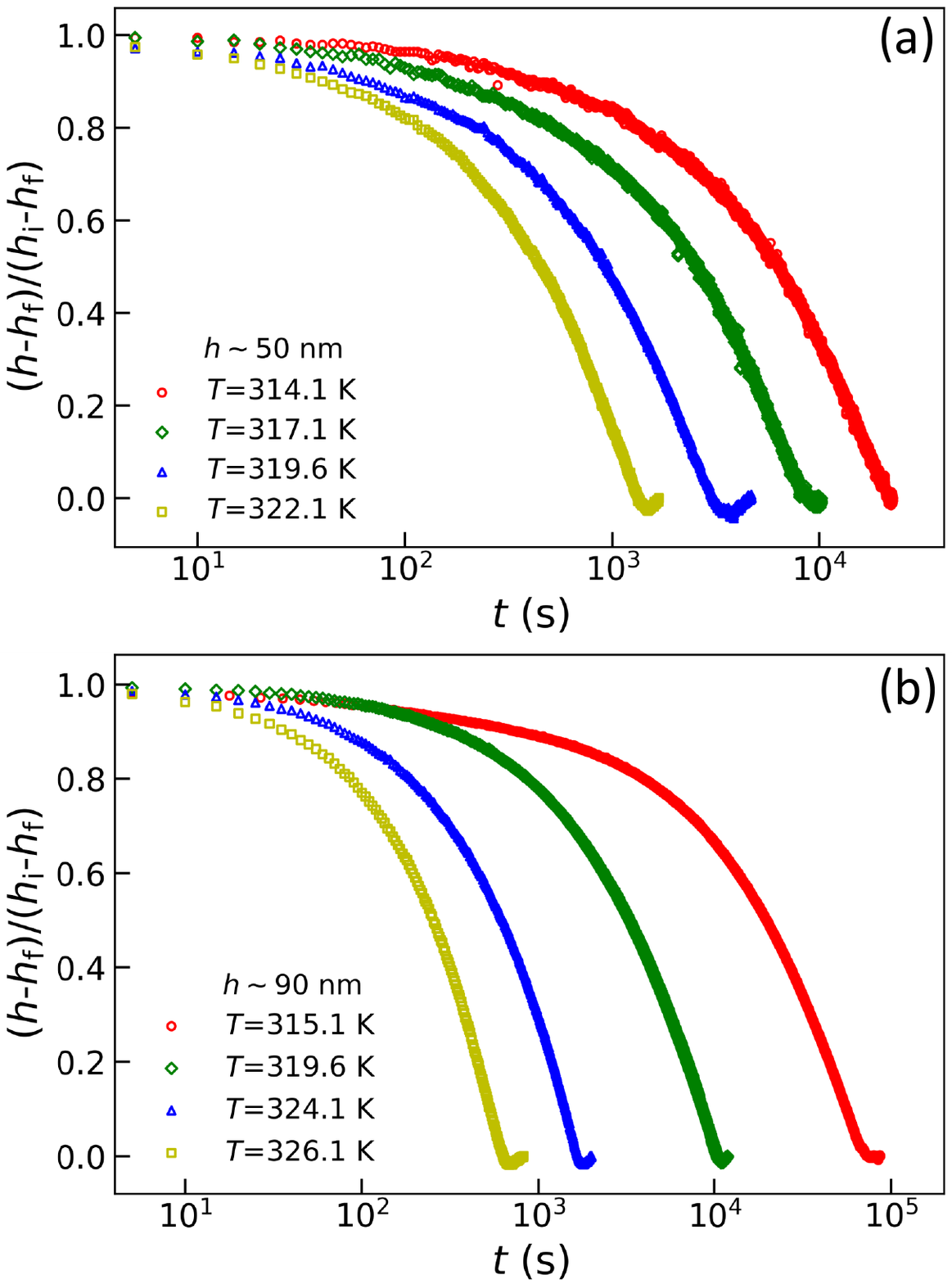}
\caption{Total film thickness $h$, normalized using the initial and final values, $h_{\textrm{i}}$ and $h_{\textrm{f}}$ respectively, as a function of time $t$ during isothermal rejuvenation, for two film thicknesses, $h \sim$ 50 nm (a) and 90 nm (b), and various temperatures $T$, as indicated.}
\label{Fig1}
\end{figure}

\section{Results}
During the rejuvenation process, there are two distinct layers in the evolving film: a supercooled-liquid layer on top of a stable-glass layer. Rejuvenation happens through the propagation of the transformation front, starting from the near-surface mobile layer and moving downwards towards the substrate~\cite{dalal2015influence,flenner2019front,rodriguez2022ultrastable,sepulveda2012stable}. As the transformation happens, the total thickness $h$ of the sample grows with time $t$. The as-prepared glasses have comparable stabilities with the $\sim90 $-nm films having a relative density increase of $\Delta \rho/\rho$ of 1.2-1.5 $ \% $, and the $\sim50$-nm films having a $\Delta \rho/\rho$ of  0.7-1.2 $ \% $.   Since we perform isothermal measurements, $\Delta \rho/\rho$ is the only measure of glass stability we can obtain for the identical samples as those used in the measurements.  While we may expect a relation between front velocity and glass stability~\cite{herrero2023front,flenner2019front}, previous results in PS showed only a weak dependence in total rejuvenation time with film stability~\cite{raegen2020ultrastable}.
\begin{figure}[h]
\centering
\includegraphics[width=\textwidth]{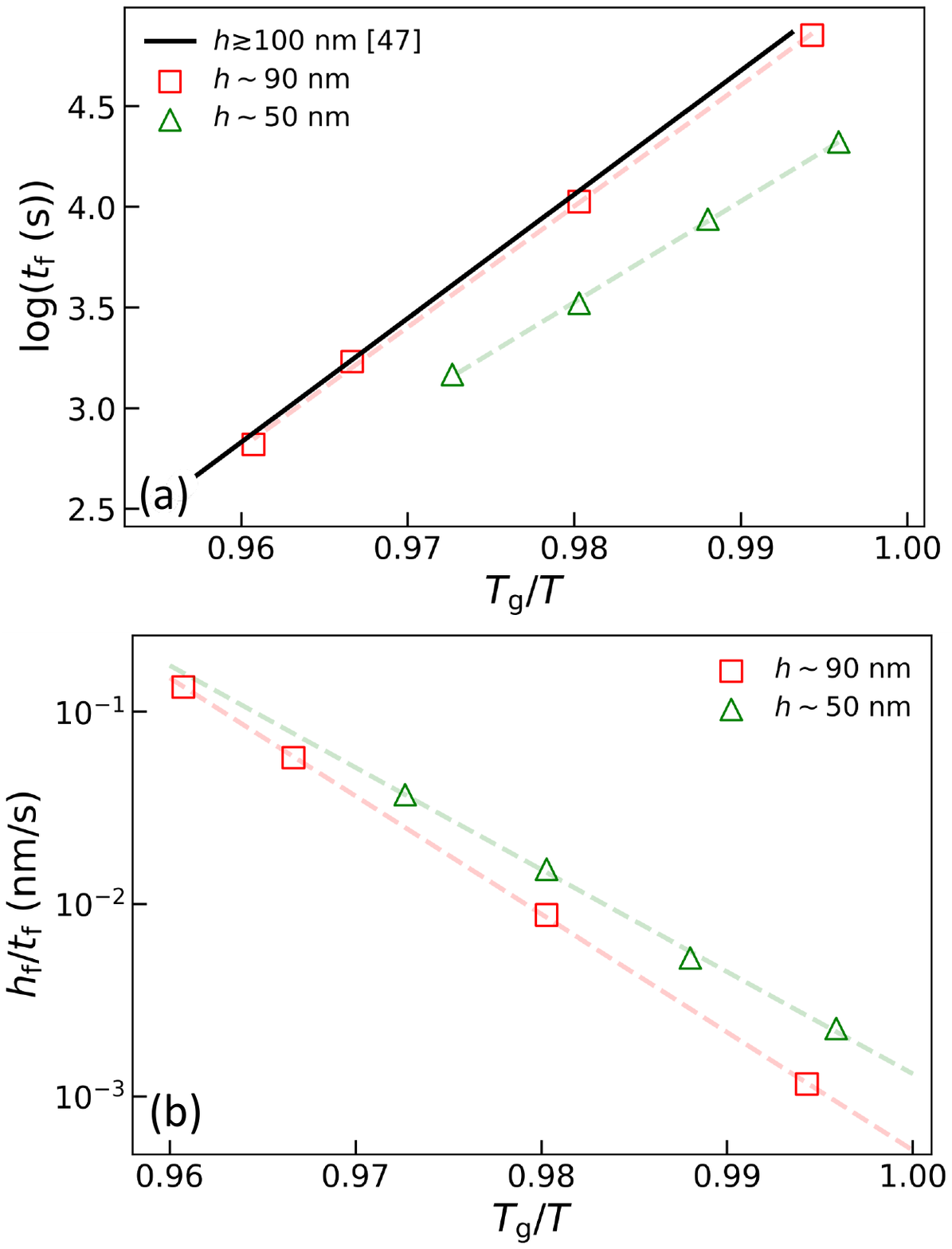}
\caption{Total time $t_{\textrm{f}}$ required for rejuvenation (a), and average front velocity $h_{\textrm{f}}/t_{\textrm{f}}$ (b), as functions of normalized inverse temperature $T_{\textrm{g}}/T$, for two different film thicknesses, as indicated. Panel (a) also shows the total rejuvenation time for the PS films of Ref.~\cite{raegen2020ultrastable}. The measurement uncertainty is represented by the symbol size. The dashed lines indicate Arrhenius-like behaviours.}
\label{Fig2}
\end{figure}

The normalized $h(t)$ data from the ellipsometric measurements are shown in Fig.~\ref{Fig1}. The most straightforward analysis is to  measure the total time $t_{\textrm{f}}$ required for the film to fully transform from the stable glass into the supercooled liquid, as a function of typical film thickness $h$ and temperature $T$. Similarly, one can extract the average velocity $h_{\textrm{f}}/t_{\textrm{f}}$ of the process, where $h_{\textrm{f}}$ is the final film thickness after rejuvenation. The results of both forms of analysis are shown in Fig.~\ref{Fig2}.  In such a small temperature range, the Vogel-Fulcher-Tammann (VFT) and Arrhenius functional forms are indistinguishable for characterising the temperature dependence of the supercooled liquid dynamics.  We employ an Arrhenius-like description of the data in Fig.~\ref{Fig2}. The slopes of the obtained linear fits in semi-log representation exhibit a clear film-thickness dependence. Moreover, we recover the same dependence with film thickness as that exhibited  by the indirect measurements of relaxation provided by the cooling-rate-dependent $T_{\textrm{g}}$ values measured in thin PS films~\cite{fakhraai2005probing}.  
\begin{figure}[h]
\centering
\includegraphics[width=\textwidth]{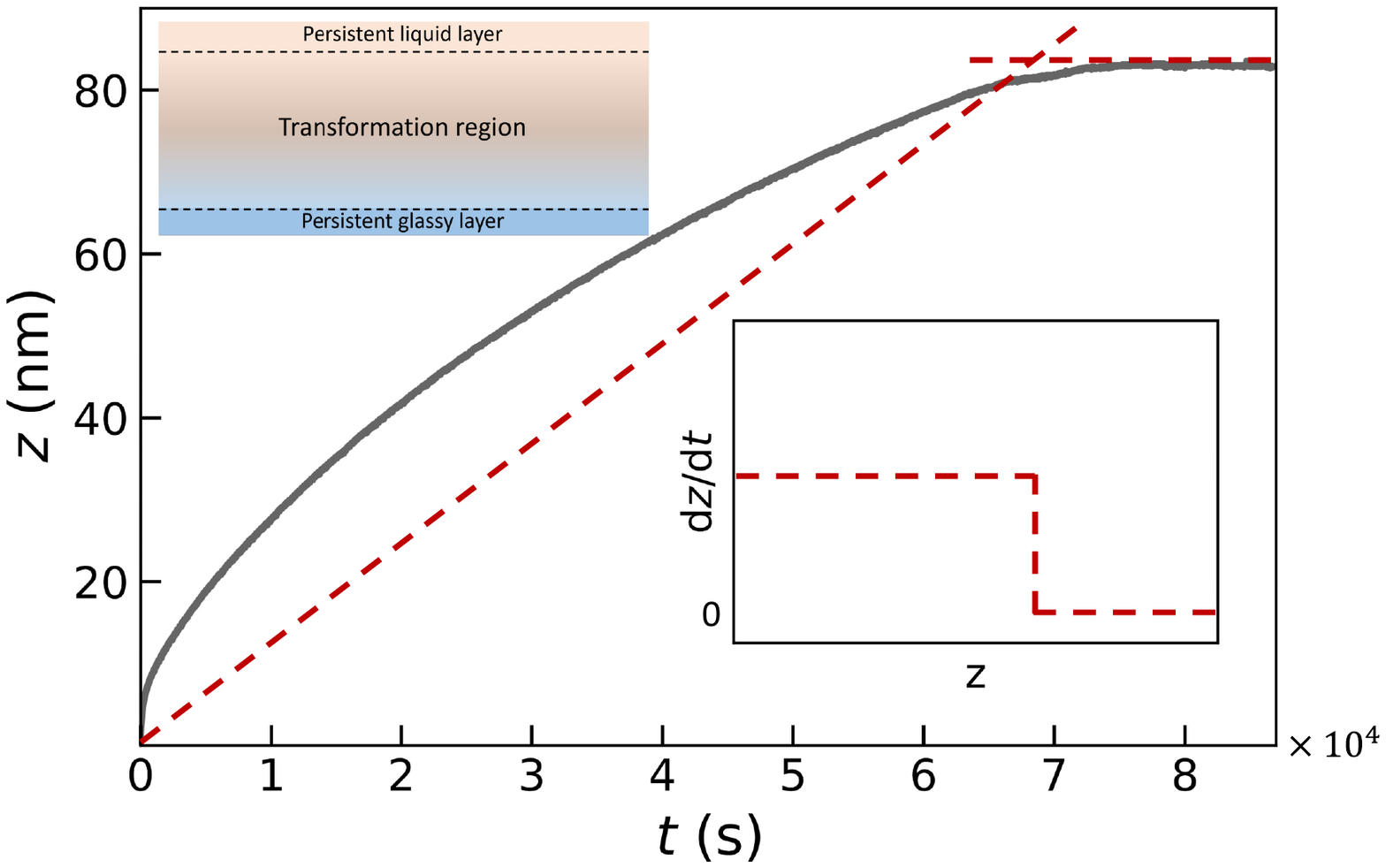}
\caption{ Depth $z$ of the rejuvenation front as a function of time $t$ (solid line), for a PS film with a typical thickness of 90~nm, at a temperature $T=315.5$~K, \textit{i.e.} 2.5~K above the measured $T_{\textrm{g}}$ of our samples. For comparison, the dashed lines indicate the expected behaviour for homogeneous dynamics~\cite{dalal2015influence,walters2015thermal}, while the lower inset shows the associated rejuvenation-front velocity $\textrm{d}z/\textrm{d}t$ versus front depth $z$ from the free surface. The upper inset shows a schematic of the  sample indicating the different film regions and notations described in the text.}
\label{Fig3}
\end{figure}

\section{Discussion}
The approach used to generate Fig.~\ref{Fig2} would provide the local relaxation dynamics only if the dynamics were homogeneous. However, we observe a clear film-thickness dependence which suggests a dynamical heterogeneity in the form of a depth-dependent mobility. We can  use the $h(t)$ data in Fig.~\ref{Fig1} to calculate the position of the rejuvenation front with time. If the initial film thickness is $h_{\textrm{i}}$ and the initial material is a stable glass at all depths, and if the final film thickness is $h_{\textrm{f}}$ and the final material is a supercooled liquid at all depths, then for a rejuvenation process which takes place through a propagating transformation front, the depth $z(t)$ of the front relative to the free surface is given at time $t$ by $z(t)= h_{\textrm{f}} \left[\frac{h(t)-h_{\textrm{i}}}{h_{\textrm{f}}-h_{\textrm{i}}}\right]$. Consequently, to extract $z(t)$, one simply needs to measure $h(t)$, which can \textit{e.g.} be provided by ellipsometry under the assumption that the refractive indices of the supercooled liquid and stable glass are the same. Comparison of initial versus final refractive index during the process reveal  a difference $\Delta n < 0.01$ between the stable glass and supercooled liquid. Because our approach follows the transformation from a stable glass to a supercooled liquid, it is naturally  insensitive to parts of the film which would not take part in such a transformation. This may include a persistent nanoscopic liquid-like layer very near the free surface~\cite{chai2020using,yin2023surface}, or, a persistent nanoscopic immobile glassy layer near the solid substrate. A schematic of these film dynamics are shown in the upper inset of  Figure~\ref{Fig3}.  References to the free surface are thus actually referring to the interface between the original stable glass and a persistent liquid layer, which has been studied and may be a few nm away from the actual free surface\cite{yin2023surface}. Figure~\ref{Fig3} shows the time dependence of the front  position $z(t)$, for a PS film with typical thickness $h \sim 90$~nm, and at a temperature $T=315.1$~K, \textit{i.e.} 2.5~K above the measured $T_{\textrm{g}}$ of our samples. Also shown for comparison in the lower inset is the dynamically homogeneous behaviour described in Refs.~\cite{dalal2015influence,walters2015thermal}, and expected for the core part of much thicker samples or for temperatures further above $T_{\textrm{g}}$. In that case, the dynamics  would be homogeneous and thus the front velocity would be constant through propagation, until the front eventually reaches the solid substrate where the velocity would change discontinuously to zero.  The measured behaviour here is qualitatively distinct from the homogeneous expectation. This  confirms the existence of some dynamical heterogeneity. More precisely, the front velocity continuously decreases as the front propagates away from the free surface and probes deeper and hence less mobile regions of the material.  While the initial change is rapid compared to the rejuvenation of the full sample, it is still much slower than thermal equilibration after the temperature jump, which allows us to discard the thermal artefacts as contributing factors. 
\begin{figure}[h]
\centering
\includegraphics[width=\textwidth]{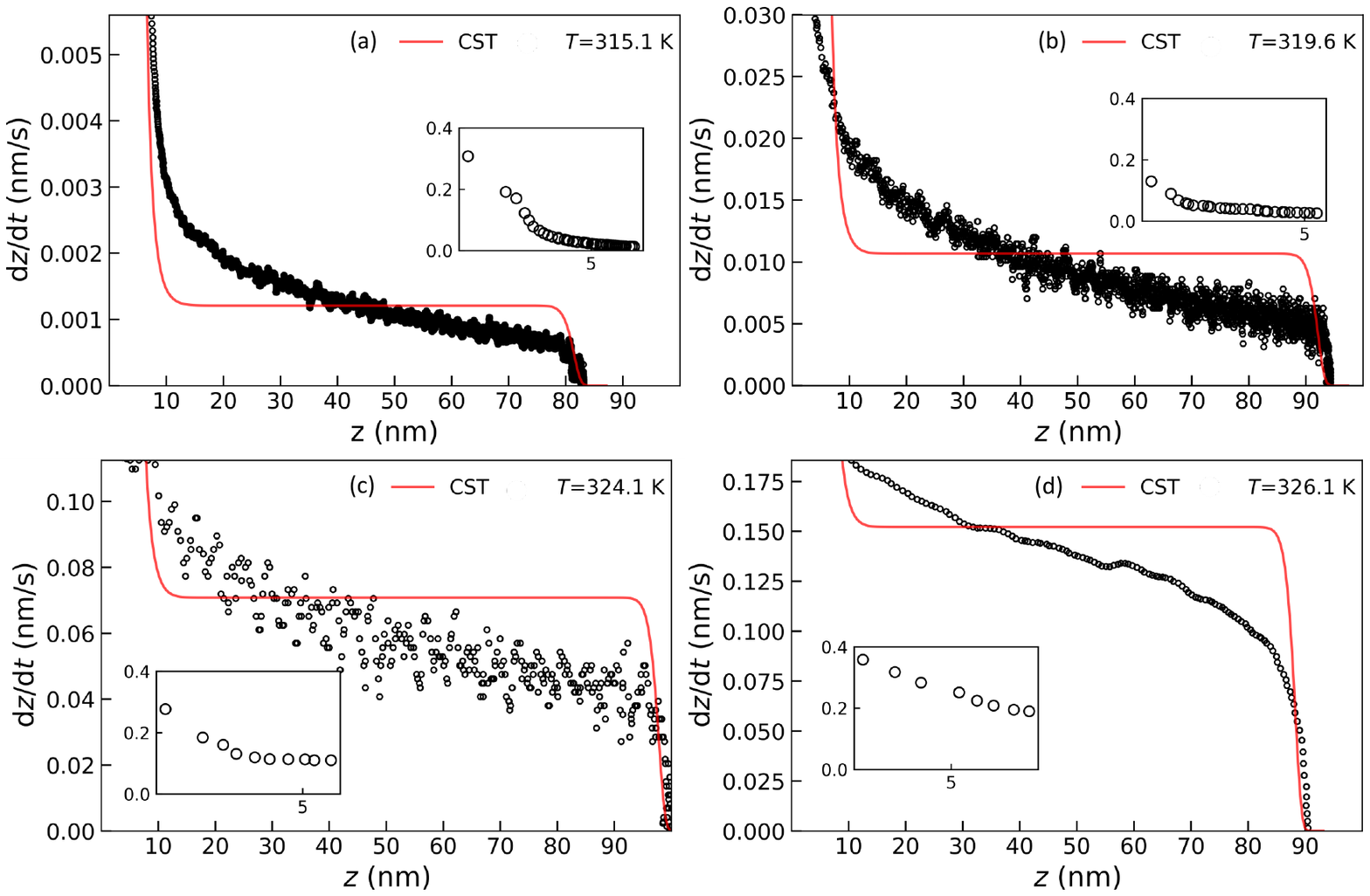}
\caption{Local rejuvenation-front velocity $\textrm{d}z/\textrm{d}t$ as a function of front depth $z$ from the position of the free surface, for various temperatures as indicated, as obtained from ellipsometric data. Insets provide zooms of the experimental data near the free surface. For comparison, predictions from the cooperative-string theory (CST) near interfaces~\cite{salez2015cooperative,arutkin2020cooperative} are shown with solid lines (see details in main text).}
\label{Fig4}
\end{figure}

The rejuvenation-front velocity $\frac{\textrm{d}z}{\textrm{d}t}$ has been correlated to the relaxation dynamics of the material adjacent to the front~\cite{rodriguez2022ultrastable}. Therefore, in order to infer the local relaxation dynamics of the supercooled material after rejuvenation, we measure the local depth-dependent front velocity from our data. We accomplish this by binning in $z$, performing linear regression in each bin, and extracting the corresponding slope. The results are shown in Fig.~\ref{Fig4}, where we plot the local front velocity $\frac{\textrm{d}z}{\textrm{d}t}$ as a function of the front depth $z$ from the free surface, for the $\sim90$-nm PS film, and for four temperatures $T$ near $T_{\textrm{g}}$. From this data, we can make three main observations. First, the interior part of the film has a front velocity which drastically increases with temperature. It varies by more than two orders of magnitude from $T=315.1$~K to $T=326.1$~K. Secondly, it is clear that the dynamics is heterogeneous, and that there is a much greater mobility near the free surface. This heterogeneity is much more pronounced at lower temperatures.  At the lowest temperature studied, the front velocity near the free surface is more than two orders of magnitude greater than that in the interior of the film. The plots allow us to immediately quantify what we mean by a near-surface region. Indeed, looking at the data in the insets, the most extreme changes in dynamics happen within the first $\sim 2-3$ nm of the free surface. Recalling that the rejuvenation-front analysis does not include the expected persistent 3-to-5-nm-thick liquid-like region in PS glassy films~\cite{chai2020using,yin2023surface},  this suggests a region of faster dynamics in PS near the free surface of about 5-8 nm in thickness at $T\approx T_{\textrm{g}}$. Interestingly, the front velocity at $z\sim0$ shows no systematic temperature dependence within the temperature range studied here. This may be influenced by our 5-s sampling time, but is also reminiscent of previous reports on the surface mobility in glassy PS~\cite{fakhraai2008measuring}, while contrasting with recent measures of the surface mobility in stable PS glasses~\cite{yin2023surface}. In any case, since the near-surface dynamics probed here has a weak temperature dependence, the increasing contribution of the near-surface region for thin films as compared to thick films naturally explains the film-thickness dependence observed in Fig.~\ref{Fig2} and in Ref.~\cite {fakhraai2005probing}. Thirdly, in all cases, the front velocity drops abruptly to zero near the solid substrate. Such a behaviour is expected to be a combination of: i) the trivial result that $\textrm{d}z/\textrm{d}t$ must drop to zero when the front hits the substrate; and ii) any effect on the supercooled-liquid dynamics due to the presence of the substrate. In the hypothetical case where the substrate has no explicit effect on the dynamics, the spatial extent of the drop to $\textrm{d}z/\textrm{d}t$=0 could be used as an in-situ evaluation of the width of the rejuvenation front -- and thus of the spatial resolution of our measurement method for the local mobility from rejuvenation-front propagation. We can then use such a worst-case scenario and the data in Fig.~\ref{Fig4} to place an upper limit on the spatial resolution of the method to be $\sim 1-2$~nm. This remarkable precision reveals that the rejuvenation front remains relatively sharp in thin films. It also allows us to rule out a parallel homogenous process, and conclude that rejuvenation is occurring only through front propagation. Computer simulations have shown that the width of the front  monotonically increases as the front moves through the film~\cite{flenner2019front,herrero2023front}, but the data in Fig.~\ref{Fig4} reveal that such a broadening is not significant for the thicknesses and temperatures studied here. 
\begin{figure}[h]
\centering
\includegraphics[width=0.8\textwidth]{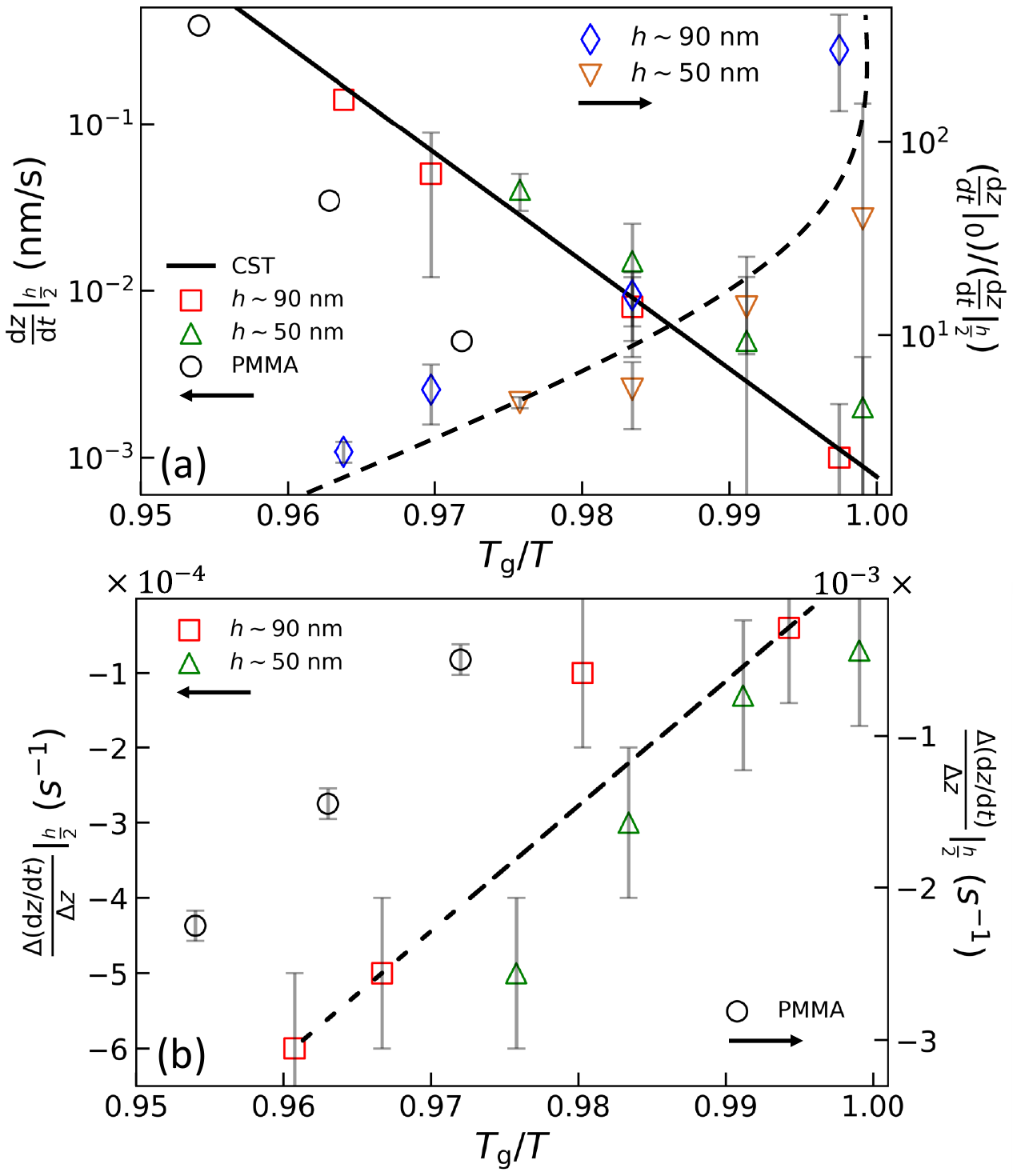}
\caption{(a) Local rejuvenation-front velocity $\textrm{d}z/\textrm{d}t$ as a function of normalized inverse temperature $T_{\textrm{g}}/T$, both: i) in the middle of the films (hollow symbols, left axis) for PS films with typical thicknesses 50~nm and 90~nm, as well as PMMA films; and ii) at the free surface (filled symbols, right axis) after normalization by the middle-point value, for PS films with typical thicknesses 50~nm and 90~nm. For comparison, the prediction from the cooperative-string theory (CST)~\cite{salez2015cooperative,arutkin2020cooperative} is shown with a solid line (see details in main text). The dashed line is a guide to the eye. (b) Local spatial derivative $\frac{\Delta(dz/dt)}{\Delta z} |_{\frac{h}{2}}$  of the rejuvenation-front velocity as a function of normalized inverse temperature $T_{\textrm{g}}/T$, for both PS films with typical thicknesses 50~nm and 90~nm (hollow symbols, left axis), and PMMA films (filled symbols, right axis). The dashed line is a guide to the eye.}
\label{Fig5}
\end{figure}

We can compare the experimental rejuvenation data above with model predictions. For this  we consider the cooperative-string theory (CST) for supercooled dynamics in confined geometries~\cite{arutkin2020cooperative}. This model is one example of minimal free-volume toy models that has the merit to capture most of the thickness-dependent glass-transition-temperature data of standard PS nanofilms in the literature~\cite{salez2015cooperative}. Specifically, we impose one free interface on one side of the film, as well as a fully-attractive interface representing the solid substrate on the other side of the film, and we calculate the depth-dependent relaxation time of a supercooled liquid at the temperature considered in each rejuvenation experiment. In practice, we separate the two interfaces and sum their independent effects, while fixing the effective film thickness from the actual $h_{\textrm{f}}$ value obtained for each sample. The model parameters are fixed as follows. The effective hard-sphere diameter $\lambda_{\textrm{V}}=3.7$~nm is set to its previously-calibrated value for PS~\cite{salez2015cooperative}. The cooperativity-onset and Vogel temperatures, $T_{\textrm{c}}=354$~K and $T_{\textrm{V}}\approx255$~K, respectively, are set so that with a molecular time scale $\tau_0=1$~ps, and a VFT barrier temperature $A=1878$~K~\cite{salez2015cooperative}, one gets a 100~s time scale for relaxation at the $T_{\textrm{g}}=313$~K value measured for the present samples. We assume an exponent $\gamma=1$~\cite{rodriguez2022ultrastable} between rejuvenation-front velocity and local relaxation rate. Finally, we invoke a single free length scale of $\sim0.6$~nm for all samples and temperatures in order to convert the predicted relaxation rate from the model to the front velocity measured in the experiments. As shown in Fig.~\ref{Fig4}, the comparison between the model and the experimental data is qualitatively good. The midpoints of the film are well described by bulk dynamics, and there is an enhancement of dynamics in the near surface region.  There is a  striking difference between the data and model predictions. For all temperatures, the experimental data shows an unexpected long-ranged depth dependence of the front velocity down to the interior region of the film, \textit{i.e.} several tens of nanometers away from the free surface. More quantitatively, the data shows a non-zero value of $\frac{\Delta (dz/dt)}{\Delta z}$.  This effect is more pronounced at higher temperatures and is in contrast with the accepted prediction (from most models), in which the deep interior of a supercooled-liquid sample should have homogeneous bulk dynamics. Recently, a numerical study has addressed in details the evolution of the front as it moves through the film~\cite{herrero2023front}. In that work, the authors have actually predicted some deviations from a homogeneous front velocity, due to the spontaneous nucleation of liquid droplets associated with the bulk rejuvenation mechanism. However, these droplets accelerate the process and lead to a broadening of the front, which is not consistent with the data shown in Fig.~\ref{Fig4}. Instead, the very large dynamical length scale experimentally observed here may perhaps be related to other similar puzzling features in polymeric glass formers~\cite{svanberg2007glass,Baglay2015,yuan2022microscale}. 

In Fig.~\ref{Fig5}(a), we plot the front-propagation velocity in the middle of the film as well as the ratio of surface to midpoint front velocities as a function of inverse temperature.  What is immediately obvious is that the  front-propagation velocity in the middle of the film is independent of film thickness within the resolution and parameter ranges of the experiments.  This confirms that the film-thickness dependence observed in Fig.~\ref{Fig2} is simply a result of the fact that the near-surface region constitutes a larger fraction of the entire sample in thinner films. The ratio of surface to midpoint front velocity shows that dynamic heterogeneity, with enhanced dynamics near the free surface, occurs over a very small range of temperatures. Similarly, Fig.~\ref{Fig5}(b) shows the anomalous variation of the front velocity with depth in the midpoint of the film (i.e. $\frac{\Delta(dz/dt)}{\Delta z} |_{\frac{h}{2}}$) as a function of $T_{\textrm{g}}/T$. From this plot we see that this anomalous dynamical process is also independent of film thickness, but that the magnitude of the effect increases with increasing temperature. The plot also shows this effect is larger in PMMA than in PS. Overall, Fig.~\ref{Fig5} indicates a lack of film-thickness dependence outside of the first few nanometers near the free surface. Conversely, any measured film-thickness dependence must come from the simple fact that thin films have a greater fraction of near-surface material than thicker films. The lack of film-thickness dependence may be the reason why layer models have proven so successful to describe film properties~\cite{forrest2014does}. 

We have included data for PMMA films in Fig.~\ref{Fig5} in order to show the applicability of the present rejuvenation-probe technique to other glass-forming materials. The technique should be applicable to any material that can be made into a stable glass in a thin-film form. It is interesting to note that Walter {\em et al.}~\cite {walters2015thermal}  also used ellipsometry to quantify front propagation in films made of an organic semiconductor, with similar thicknesses. In that work, the authors were able to model the ellipsometric data with a layer model and to identify the front even more directly. Despite many similarities to the current work, they saw no evidence for enhanced mobility near the free surface. Such a difference in behaviours may be partially attributed to a difference in mobility gradients for different materials.  Additionally, the results are drastically impacted by the temperatures probed in the experiments. As shown in Fig.~\ref{Fig5}, the dynamical heterogeneity (measured by the ratio of maximum front velocity to the value at the midpoint of the film) varies from $\sim 300$ at $T_{\textrm{g}}/T \approx 1$ to $\sim 2$ at $T_{\textrm{g}}/T=0.95$. These values are consistent with the onset of anomalous surface behaviours happening only at temperatures very near $T_{\textrm{g}}$~\cite{fakhraai2008measuring,fakhraai2005probing,chai2014direct}. In Ref.~\cite{walters2015thermal}, the lowest temperature $T$ where isothermal rejuvenation is measured corresponds to $T_{\textrm{g}}/T=0.97$.   This combination of different materials, and narrow temperature range of expected observation may  account for the difference between the respective experiments' abilities to detect  enhanced mobility in the near-surface region.

In summary, we have studied the rejuvenation of nanofilms of stable-glassy polystyrene at temperatures near and above their glass-transition temperature. We have demonstrated that we can use the mechanism of rejuvenation by front propagation as a highly-localized nanorheological probe of the supercooled-liquid dynamics through the film. This provides a novel route to quantify the depth-dependent relaxation dynamics in glass-forming films -- and even in relatively thick samples -- with unprecedented precision. The results show a significant enhancement of the local dynamics in the first few nanometers near the liquid-like surface layer at temperatures very near $T_{\textrm{g}}$. The temperature dependence of the dynamics deep in the interior of the film shows no film-thickness dependence and is well described by bulk temperature dependence -- showing that film-thickness dependences in such samples simply originate from the volume fraction of the near-surface material. Finally, an extra and unexpected long-ranged depth dependence of the supercooled dynamics was observed, with no dependence on film thickness. The cause of this extra relaxation process is unknown and should be  studied in future investigations.  

The authors thank Mark Ediger for interesting discussions and comments. The authors acknowledge financial support from the Natural Sciences and Engineering Research council of Canada (NSERC), as well as from the European Union through the European Research Council under EMetBrown (ERC-CoG-101039103) grant. Views and opinions expressed are however those of the authors only and do not necessarily reflect those of the European Union or the European Research Council. Neither the European Union nor the granting authority can be held responsible for them. The authors also acknowledge financial support from the Agence Nationale de la Recherche under EMetBrown (ANR-21-ERCC-0010-01), Softer (ANR-21-CE06-0029), and Fricolas (ANR-21-CE06-0039) grants. Finally, JAF is grateful to the University of Bordeaux for funding provided through the visiting scholar program. 

\section{Materials and methods}
Films of stable polystyrene (PS) glass are prepared by vapour deposition of monodisperse PS materials onto a substrate held at a temperature below the $T_{\textrm{g}}$ value (obtained by ellipsometry) of that sample, as described previously~\cite{raegen2020ultrastable}. Two sets of samples are studied. One set, with a typical thickness of $\sim 90$~nm, has thicknesses in the 83-98~nm range; and the other, with a typical thickness of $\sim 50$~nm, has thicknesses in the 45-55~nm range. The $T_{\textrm{g}}$ values of all samples, as measured by cooling the equilibrium liquid at 2K/min, are $T_{\textrm{g}}=313 \pm 1$~K. After preparation, the samples are kept at 277~K until ready for use (10-12 days for these samples).  Samples are then placed on a Linkam hot stage and ellipsometric data is collected by a Film Sense  FS-1EX ellipsometer at a rate of a point per 5 seconds. The sample temperature is ramped to the desired value at a rate of 5 K/min. This value is chosen as a compromise between reaching the measurement temperature as quickly as possible, but without any overshoot. The ellipsometric data is analyzed using the instrument software.  We note that fitting our ellipsometric data with a multilayer model, where the supercooled liquid and stable glass have different refractive indices~~\cite{dalal2015influence,walters2015thermal}, does not provide a better fit than the approximation of a single layer where the film is modelled as a homogeneous material of time-dependent thickness $h(t)$ and refractive index $n(t)$.  For this reason, a single-layer model is used to represent the ellipsometric data. The same approximation is used for each temperature, and so, for any given film thickness, the comparisons between different temperatures are robust. The difference between the success of single-layer modelling here versus the multilayer approach successful in Ref.~\cite{walters2015thermal} may be at least partially due to the lack of birefringence in the PS stable glass~\cite{raegen2020ultrastable}, which results in much less contrast at the liquid-glass interface. PMMA samples with a thickness range 96-148 nm  were also prepared and characterised in the same way as for PS.
\bibliographystyle{ieeetr}
\bibliography{Karimi2023}

\begin{thebibliography}{10}

\bibitem{keddie1994size}
J.~L. Keddie, R.~A. Jones, and R.~A. Cory, ``Size-dependent depression of the
  glass transition temperature in polymer films,'' {\em EPL (Europhysics
  Letters)}, vol.~27, no.~1, p.~59, 1994.

\bibitem{roth2005glass}
C.~B. Roth and J.~R. Dutcher, ``Glass transition and chain mobility in thin
  polymer films,'' {\em Journal of Electroanalytical Chemistry}, vol.~584,
  no.~1, pp.~13--22, 2005.

\bibitem{forrest2001glass}
J.~A. Forrest and K.~Dalnoki-Veress, ``The glass transition in thin polymer
  films,'' {\em Advances in Colloid and Interface Science}, vol.~94, no.~1-3,
  pp.~167--195, 2001.

\bibitem{ediger2014dynamics}
M.~D. Ediger and J.~A. Forrest, ``Dynamics near free surfaces and the glass
  transition in thin polymer films: a view to the future,'' {\em
  Macromolecules}, vol.~47, no.~2, pp.~471--478, 2014.

\bibitem{alcoutlabi2005effects}
M.~Alcoutlabi and G.~B. McKenna, ``Effects of confinement on material behaviour
  at the nanometre size scale,'' {\em Journal of Physics: Condensed Matter},
  vol.~17, no.~15, p.~R461, 2005.

\bibitem{fujimoto2020energy}
D.~Fujimoto, W.~A. MacFarlane, and J.~Rottler, ``Energy barriers and
  cooperative motion at the surface of freestanding glassy polystyrene films,''
  {\em The Journal of Chemical Physics}, vol.~153, no.~15, 2020.

\bibitem{baljon2010simulated}
A.~Baljon, S.~Williams, N.~Balabaev, F.~Paans, D.~Hudzinskyy, and A.~Lyulin,
  ``Simulated glass transition in free-standing thin polystyrene films,'' {\em
  Journal of Polymer Science Part B: Polymer Physics}, vol.~48, no.~11,
  pp.~1160--1167, 2010.

\bibitem{hao2021mobility}
Z.~Hao, A.~Ghanekarade, N.~Zhu, K.~Randazzo, D.~Kawaguchi, K.~Tanaka, X.~Wang,
  D.~S. Simmons, R.~D. Priestley, and B.~Zuo, ``Mobility gradients yield
  rubbery surfaces on top of polymer glasses,'' {\em Nature}, vol.~596,
  no.~7872, pp.~372--376, 2021.

\bibitem{ivancic2020dynamic}
R.~J. Ivancic and R.~A. Riggleman, ``Dynamic phase transitions in freestanding
  polymer thin films,'' {\em Proceedings of the National Academy of Sciences},
  vol.~117, no.~41, pp.~25407--25413, 2020.

\bibitem{torres2000molecular}
J.~Torres, P.~Nealey, and J.~J. de~Pablo, ``Molecular simulation of ultrathin
  polymeric films near the glass transition,'' {\em Physical Review Letters},
  vol.~85, no.~15, p.~3221, 2000.

\bibitem{zhou2017short}
Y.~Zhou and S.~T. Milner, ``Short-time dynamics reveals t g suppression in
  simulated polystyrene thin films,'' {\em Macromolecules}, vol.~50, no.~14,
  pp.~5599--5610, 2017.

\bibitem{forrest2014does}
J.~A. Forrest and K.~Dalnoki-Veress, ``When does a glass transition temperature
  not signify a glass transition?,'' {\em ACS Macro Letters}, vol.~3, no.~4,
  pp.~310--314, 2014.

\bibitem{lipson2010local}
J.~E. Lipson and S.~T. Milner, ``Local and average glass transitions in polymer
  thin films,'' {\em Macromolecules}, vol.~43, no.~23, pp.~9874--9880, 2010.

\bibitem{salez2015cooperative}
T.~Salez, J.~Salez, K.~Dalnoki-Veress, E.~Rapha{\"e}l, and J.~A. Forrest,
  ``Cooperative strings and glassy interfaces,'' {\em Proceedings of the
  National Academy of Sciences}, vol.~112, no.~27, pp.~8227--8231, 2015.

\bibitem{zhang2018we}
W.~Zhang, J.~F. Douglas, and F.~W. Starr, ``Why we need to look beyond the
  glass transition temperature to characterize the dynamics of thin supported
  polymer films,'' {\em Proceedings of the National Academy of Sciences},
  vol.~115, no.~22, pp.~5641--5646, 2018.

\bibitem{forrest1998relaxation}
J.~Forrest, C.~Svanberg, K.~Revesz, M.~Rodahl, L.~Torell, and B.~Kasemo,
  ``Relaxation dynamics in ultrathin polymer films,'' {\em Physical Review E},
  vol.~58, no.~2, p.~R1226, 1998.

\bibitem{svanberg2007glass}
C.~Svanberg, ``Glass transition relaxations in thin suspended polymer films,''
  {\em Macromolecules}, vol.~40, no.~2, pp.~312--315, 2007.

\bibitem{yuan2022microscale}
H.~Yuan, J.~Yan, P.~Gao, S.~K. Kumar, and O.~K. Tsui, ``Microscale mobile
  surface double layer in a glassy polymer,'' {\em Science Advances}, vol.~8,
  no.~45, p.~eabq5295, 2022.

\bibitem{o2005rheological}
P.~O'connell and G.~McKenna, ``Rheological measurements of the
  thermoviscoelastic response of ultrathin polymer films,'' {\em Science},
  vol.~307, no.~5716, pp.~1760--1763, 2005.

\bibitem{fakhraai2005probing}
Z.~Fakhraai and J.~A. Forrest, ``Probing slow dynamics in supported thin
  polymer films,'' {\em Physical Review Letters}, vol.~95, no.~2, p.~025701,
  2005.

\bibitem{koh2008structural}
Y.~P. Koh and S.~L. Simon, ``Structural relaxation of stacked ultrathin
  polystyrene films,'' {\em Journal of Polymer Science Part B: Polymer
  Physics}, vol.~46, no.~24, pp.~2741--2753, 2008.

\bibitem{tian2022surface}
H.~Tian, Q.~Xu, H.~Zhang, R.~D. Priestley, and B.~Zuo, ``Surface dynamics of
  glasses,'' {\em Applied Physics Reviews}, vol.~9, no.~1, 2022.

\bibitem{sharp2003free}
J.~S. Sharp and J.~A. Forrest, ``Free surfaces cause reductions in the glass
  transition temperature of thin polystyrene films,'' {\em Physical Review
  Letters}, vol.~91, no.~23, p.~235701, 2003.

\bibitem{chai2014direct}
Y.~Chai, T.~Salez, J.~D. McGraw, M.~Benzaquen, K.~Dalnoki-Veress,
  E.~Rapha{\"e}l, and J.~A. Forrest, ``A direct quantitative measure of surface
  mobility in a glassy polymer,'' {\em Science}, vol.~343, no.~6174,
  pp.~994--999, 2014.

\bibitem{daley2012comparing}
C.~Daley, Z.~Fakhraai, M.~Ediger, and J.~Forrest, ``Comparing surface and bulk
  flow of a molecular glass former,'' {\em Soft Matter}, vol.~8, no.~7,
  pp.~2206--2212, 2012.

\bibitem{zhang2017decoupling}
Y.~Zhang and Z.~Fakhraai, ``Decoupling of surface diffusion and relaxation
  dynamics of molecular glasses,'' {\em Proceedings of the National Academy of
  Sciences}, vol.~114, no.~19, pp.~4915--4919, 2017.

\bibitem{li2022surface}
Y.~Li, C.~Bishop, K.~Cui, J.~Schmidt, M.~Ediger, and L.~Yu, ``Surface diffusion
  of a glassy discotic organic semiconductor and the surface mobility gradient
  of molecular glasses,'' {\em The Journal of Chemical Physics}, vol.~156,
  no.~9, 2022.

\bibitem{ellison2003distribution}
C.~J. Ellison and J.~M. Torkelson, ``The distribution of glass-transition
  temperatures in nanoscopically confined glass formers,'' {\em Nature
  materials}, vol.~2, no.~10, pp.~695--700, 2003.

\bibitem{priestley2005structural}
R.~D. Priestley, C.~J. Ellison, L.~J. Broadbelt, and J.~M. Torkelson,
  ``Structural relaxation of polymer glasses at surfaces, interfaces, and in
  between,'' {\em Science}, vol.~309, no.~5733, pp.~456--459, 2005.

\bibitem{mckenzie2015enhanced}
I.~McKenzie, C.~R. Daley, R.~F. Kiefl, C.~P. Levy, W.~A. MacFarlane, G.~D.
  Morris, M.~R. Pearson, D.~Wang, and J.~A. Forrest, ``Enhanced high-frequency
  molecular dynamics in the near-surface region of polystyrene thin films
  observed with $\beta$-nmr,'' {\em Soft Matter}, vol.~11, no.~9,
  pp.~1755--1761, 2015.

\bibitem{mckenzie2022beta}
I.~McKenzie, D.~Fujimoto, V.~L. Karner, R.~Li, W.~A. MacFarlane, R.~M.
  McFadden, G.~D. Morris, M.~R. Pearson, A.~N. Raegen, M.~Stachura, {\em
  et~al.}, ``A $\beta$-nmr study of the depth, temperature, and
  molecular-weight dependence of secondary dynamics in polystyrene:
  Entropy--enthalpy compensation and dynamic gradients near the free surface,''
  {\em The Journal of Chemical Physics}, vol.~156, no.~8, 2022.

\bibitem{oba2012relaxation}
T.~Oba and M.~Vacha, ``Relaxation in thin polymer films mapped across the film
  thickness by astigmatic single-molecule imaging,'' {\em ACS Macro Letters},
  vol.~1, no.~6, pp.~784--788, 2012.

\bibitem{swallen2007organic}
S.~F. Swallen, K.~L. Kearns, M.~K. Mapes, Y.~S. Kim, R.~J. McMahon, M.~D.
  Ediger, T.~Wu, L.~Yu, and S.~Satija, ``Organic glasses with exceptional
  thermodynamic and kinetic stability,'' {\em Science}, vol.~315, no.~5810,
  pp.~353--356, 2007.

\bibitem{rodriguez2022ultrastable}
C.~Rodriguez-Tinoco, M.~Gonzalez-Silveira, M.~A. Ramos, and J.~Rodriguez-Viejo,
  ``Ultrastable glasses: new perspectives for an old problem,'' {\em La Rivista
  del Nuovo Cimento}, vol.~45, no.~5, pp.~325--406, 2022.

\bibitem{ediger2017perspective}
M.~D. Ediger, ``Perspective: Highly stable vapor-deposited glasses,'' {\em The
  Journal of Chemical Physics}, vol.~147, no.~21, p.~210901, 2017.

\bibitem{parmar2020ultrastable}
A.~D. Parmar, M.~Ozawa, and L.~Berthier, ``Ultrastable metallic glasses in
  silico,'' {\em Physical Review Letters}, vol.~125, no.~8, p.~085505, 2020.

\bibitem{berthier2020measure}
L.~Berthier and M.~D. Ediger, ``How to measure a structural relaxation time
  that is too long to be measured?,'' {\em The Journal of Chemical Physics},
  vol.~153, no.~4, 2020.

\bibitem{lyubimov2013model}
I.~Lyubimov, M.~D. Ediger, and J.~J. de~Pablo, ``Model vapor-deposited glasses:
  Growth front and composition effects,'' {\em The Journal of chemical
  physics}, vol.~139, no.~14, 2013.

\bibitem{lin2014molecular}
P.-H. Lin, I.~Lyubimov, L.~Yu, M.~Ediger, and J.~J. de~Pablo, ``Molecular
  modeling of vapor-deposited polymer glasses,'' {\em The Journal of chemical
  physics}, vol.~140, no.~20, 2014.

\bibitem{berthier2017origin}
L.~Berthier, P.~Charbonneau, E.~Flenner, and F.~Zamponi, ``Origin of
  ultrastability in vapor-deposited glasses,'' {\em Physical review letters},
  vol.~119, no.~18, p.~188002, 2017.

\bibitem{flenner2019front}
E.~Flenner, L.~Berthier, P.~Charbonneau, and C.~J. Fullerton, ``Front-mediated
  melting of isotropic ultrastable glasses,'' {\em Physical review letters},
  vol.~123, no.~17, p.~175501, 2019.

\bibitem{swallen2009stable}
S.~F. Swallen, K.~Traynor, R.~J. McMahon, M.~Ediger, and T.~E. Mates, ``Stable
  glass transformation to supercooled liquid via surface-initiated growth
  front,'' {\em Physical review letters}, vol.~102, no.~6, p.~065503, 2009.

\bibitem{dalal2015influence}
S.~S. Dalal and M.~D. Ediger, ``Influence of substrate temperature on the
  transformation front velocities that determine thermal stability of
  vapor-deposited glasses,'' {\em The Journal of Physical Chemistry B},
  vol.~119, no.~9, pp.~3875--3882, 2015.

\bibitem{walters2015thermal}
D.~M. Walters, R.~Richert, and M.~D. Ediger, ``Thermal stability of
  vapor-deposited stable glasses of an organic semiconductor,'' {\em The
  Journal of Chemical Physics}, vol.~142, no.~13, 2015.

\bibitem{sepulveda2012stable}
A.~Sep{\'u}lveda, S.~F. Swallen, L.~A. Kopff, R.~J. McMahon, and M.~Ediger,
  ``Stable glasses of indomethacin and $\alpha$, $\alpha$,
  $\beta$-tris-naphthylbenzene transform into ordinary supercooled liquids,''
  {\em The Journal of chemical physics}, vol.~137, no.~20, 2012.

\bibitem{herrero2023front}
C.~Herrero, M.~D. Ediger, and L.~Berthier, ``Front propagation in ultrastable
  glasses is dynamically heterogeneous,'' {\em Journal of Chemical Physics},
  vol.~159, no.~11, p.~114504, 2023.

\bibitem{raegen2020ultrastable}
A.~N. Raegen, J.~Yin, Q.~Zhou, and J.~A. Forrest, ``Ultrastable monodisperse
  polymer glass formed by physical vapour deposition,'' {\em Nature Materials},
  vol.~19, no.~10, pp.~1110--1113, 2020.

\bibitem{chai2020using}
Y.~Chai, T.~Salez, and J.~A. Forrest, ``Using $m_w$ dependence of surface
  dynamics of glassy polymers to probe the length scale of free-surface
  mobility,'' {\em Macromolecules}, vol.~53, no.~3, pp.~1084--1089, 2020.

\bibitem{yin2023surface}
J.~Yin, C.~Pedersen, M.~F. Thees, A.~Carlson, T.~Salez, and J.~A. Forrest,
  ``Surface and bulk relaxation of vapor-deposited polystyrene glasses,'' {\em
  The Journal of Chemical Physics}, vol.~158, no.~9, 2023.

\bibitem{arutkin2020cooperative}
M.~Arutkin, E.~Rapha{\"e}l, J.~A. Forrest, and T.~Salez, ``Cooperative strings
  and glassy dynamics in various confined geometries,'' {\em Physical Review
  E}, vol.~101, no.~3, p.~032122, 2020.

\bibitem{fakhraai2008measuring}
Z.~Fakhraai and J.~Forrest, ``Measuring the surface dynamics of glassy
  polymers,'' {\em Science}, vol.~319, no.~5863, pp.~600--604, 2008.

\bibitem{Baglay2015}
R.~R. Baglay and C.~B. Roth, ``{Experimentally determined profile of local
  glass transition temperature across a glassy-rubbery polymer interface with a
  Tg difference of 80 K},'' {\em The Journal of Chemical Physics}, vol.~143,
  no.~11, p.~111101, 2015.

\end{thebibliography}
\end{document}